# Catalog of Ultraviolet Bright Stars (CUBS): Strategies for UV occultation measurements, planetary illumination modeling, and sky map analyses using hybrid IUE-Kurucz spectra

Michael A. Velez[1,2,*], Kurt D. Retherford[2,1], Vincent Hue[3], Joshua A. Kammer[2], Tracy M. Becker[2,1], G. Randall Gladstone[2,1], Michael W. Davis[2], Thomas K. Greathouse[2], Philippa M. Molyneux[2], Shawn M. Brooks[3], Ujjwal Raut[2,1], Maarten H. Versteeg[2]

[1]Department of Physics and Astronomy, University of Texas at San Antonio, 1 UTSA Circle, San Antonio, TX, USA 78249
[2]Southwest Research Institute, 6220 Culebra Road, San Antonio, TX, USA 78238
[3]Laboratoire d'Astrophysique de Marseille, 38 Rue Frédéric Joliot Curie, 13013 Marseille, France
[4]Jet Propulsion Laboratory, California Institute of Technology, 4800 Oak Grove Drive, Pasadena, CA, USA 91109

**Abstract**. Ultraviolet spectroscopy is a powerful method to study planetary surface composition through reflectance measurements and atmospheric composition through stellar/solar occultations, transits of other planetary bodies, and direct imaging of airglow and auroral emissions. The next generation of ultraviolet spectrographs (UVS) on board ESA's JUICE (Jupiter Icy Moons Explorer) and NASA's Europa Clipper missions will perform such measurements of Jupiter and its moons in the early 2030's. This work presents a compilation of a detailed UV stellar catalog, named CUBS, of targets with high intensity in the 50-210 nm wavelength range with applications relevant to planetary spectroscopy. These applications include: 1) Planning and simulating occultations, including calibration measurements; 2) Modeling starlight illumination of dark, nightside planetary surfaces primarily lit by the sky; and 3) Studying the origin of diffuse galactic UV light as mapped by existing datasets from Juno-UVS and others. CUBS includes information drawn from resources such as the International Ultraviolet Explorer (IUE) catalog and SIMBAD. We have constructed model spectra at 0.1 nm resolution for almost 90,000 targets using Kurucz models and, when available, IUE spectra. CUBS also includes robust checks for agreement between the Kurucz models and the IUE data. We also present a tool for which our catalog can be used to identify the best candidates for stellar occultation observations, with applications for any UV instrument. We report on our methods for producing CUBS and discuss plans for its implementation during ongoing and upcoming planetary missions.

## 1 INTRODUCTION

Ultraviolet spectroscopy is an important tool for space science. Spectroscopic techniques involving the use of UV-bright stars include, but are not limited to: stellar occultation/appulse observations for atmospheric studies, measurements of dark or shadowed surfaces through starlight illumination, and investigating diffuse UV galactic dust emissions. These applications motivated the creation of a stellar catalog of UV-bright stars in order to provide simpler ways to plan and execute observations that invoke these techniques. One of the primary motivations for compiling this catalog is NASA's upcoming Europa Clipper mission[1], which is set to perform transformational science focusing on Europa, Jupiter's second closest Galilean moon. The Europa Ultraviolet Spectrograph (Europa-UVS) will conduct a series of observations, including those mentioned above, that require knowledge about the stars in the UV. Additionally, ESA's JUICE mission (Jupiter Icy Moons Explorer) will carry a UV instrument, JUICE-UVS, from the same generation as Europa-UVS. JUICE-UVS will explore the Jupiter system and its moons.

UV spectrographs on previous space missions have used these observational techniques, and Europa-UVS is the latest in a series of spectrographs that will complete significant UV observations throughout the solar system. The Alice instrument on ESA's Rosetta mission was the first UV spectrograph to study a comet (67P/Churyumov-Gerasimenko) at close range[2]. It used stellar appulse observations to study $H_2O$ and $O_2$ in the coma of the comet[3]. The Alice instrument on NASA's New Horizons mission was the first to closely study Pluto, Charon, and Arrokoth[4]. During its cruise to Pluto, New Horizons' Alice also observed a stellar occultation by Jupiter's atmosphere which helped to constrain the concentration of compounds such as methane, acetylene, and ethane[5]. At Pluto, New Horizons' Alice observed solar and stellar occultations by Pluto's atmosphere and constrained profiles for nitrogen as well as hydrocarbon compounds[6,7,8]. The LAMP (Lyman Alpha Mapping Project) instrument on NASA's LRO (Lunar Reconnaissance Orbiter) continues to monitor the Moon's surface and atmosphere[9]. One of its main functions is to analyze permanently shadowed regions on the surface using starlight illumination rather than sunlight[10,11]. Juno-UVS on NASA's Juno mission studies far-UV auroral emissions at Jupiter[12]. Notably, Juno-UVS has obtained a full spectral map of the sky in the UV that is considered in our study[13,14]. Both the upcoming Europa-UVS and JUICE-UVS instruments will also be able to conduct studies of the Jovian system using these techniques. Other important UV instruments include Cassini UVIS (Ultraviolet Imaging Spectrograph)[15] and the SPICAM (Spectroscopy for the investigation of the characteristics of the atmosphere of Mars) Light UV spectrometer on Mars Express[16]. Cassini UVIS performed stellar and solar occultations by Saturn[17,18] and its rings[19,20] as well as by the water vapor plumes of Enceladus[21,22]. SPICAM was able to detect molecular oxygen on Mars using stellar occultation measurements of the planet[23] and further demonstrate how far-UV observations are effective for the study of planetary bodies and the galaxy as a whole.

*michael.velez@contractor.swri.org

The CUBS catalog initially began as a much smaller endeavor with 1,000 stars with available spectra from the International Ultraviolet Explorer (IUE). The catalog was first used to model starlight illumination of the Moon's surface for LAMP observations[11] (Section 3.2) and was later expanded to 2,000 stars by adding other UV-bright stars that could be well-represented by Kurucz models[11]. A separate catalog of Juno-UVS spectra of >500 stellar targets obtained to perform instrument calibrations[14] provides a valuable cross-comparison (as reported here) and identified a few targets that had been missing from the original catalog. This paper will discuss the methods used to create this catalog (Section 2), identify three planned direct applications for this catalog (Section 3), and summarize the conclusions of this work, most notably how well the Kurucz models perform when compared to observations (Section 4).

## 2 METHODOLOGY

CUBS currently consists of nearly 90,000 stars of O (~5,000), B (~41,000), and A (~42,000) types. The entirety of CUBS exists as a spreadsheet and an IDL save file. The catalog spreadsheet contains specific information about each star (where available) retrieved from the SIMBAD database[24]. These data include: HD Number, Common Identifier, Right Ascension, Declination, Spectral Type, Stellar Type, B Magnitude, V Magnitude, Proper Motion (both RA and Dec), Radial Velocity, and Parallax. All numerical quantities are included with their uncertainties. The IDL save file contains modeled spectral flux density of each star in CUBS within the wavelength range of 50-210 nm (JUICE-UVS covers a range of 50-204 nm and Europa-UVS similarly covers 55-206 nm). These spectra were produced using two different methods: Kurucz models alone or a combination of IUE data and Kurucz models.

Kurucz models were developed originally in 1979[25] and were updated in 1992[26] with new data and improved modeling resolution, such as a higher number of optical depth layers. The updated

Kurucz Atlas contains roughly 7600 stellar atmosphere models with wavelengths ranging from 9 nm to 160,000 nm at a spectral resolution of 1 nm, which helps ensure the robustness of CUBS spectra for all stellar types. Castelli and Kurucz[27] updated the stellar model to include interstellar absorptions from 140-160 nm, among other improvements. To construct the spectra, we use key parameters for each star such as its stellar type to get an applicable effective temperature that can be used as an input to the Kurucz model. The B and V magnitudes for each star are then applied to redden the spectra as they would be observed from the solar system.

These Kurucz models are supplemented by observational data from the IUE where available. The IUE was launched in 1978 to observe UV-bright stars, specifically within the range of 115 to 320 nm with spectral resolutions ranging from 0.02 nm to 0.6 nm[28]. Throughout the entire duration of the mission, IUE observed over 100,000 targets of which around 1,800 have spectra that were deemed reliable to use for Juno-UVS calibrations and are also used in our catalog[14]. These IUE spectra cover the wavelength range of 116-196 nm, and so need to be extended out to the full range of CUBS by normalizing the Kurucz spectra to the IUE data via interpolation and appending it to the short-end (50-116 nm) and long-end (196-210 nm) wavelength regions for UVS. For the purposes of modeling the estimated signal measured by a specific UV instrument, the stellar flux can be converted to an instrument response using its effective area curve, i.e., the effective area of the telescope mirror after accounting for the throughput losses. Figure 1 shows an example of both a flux spectrum and an instrument count spectrum for Europa-UVS.

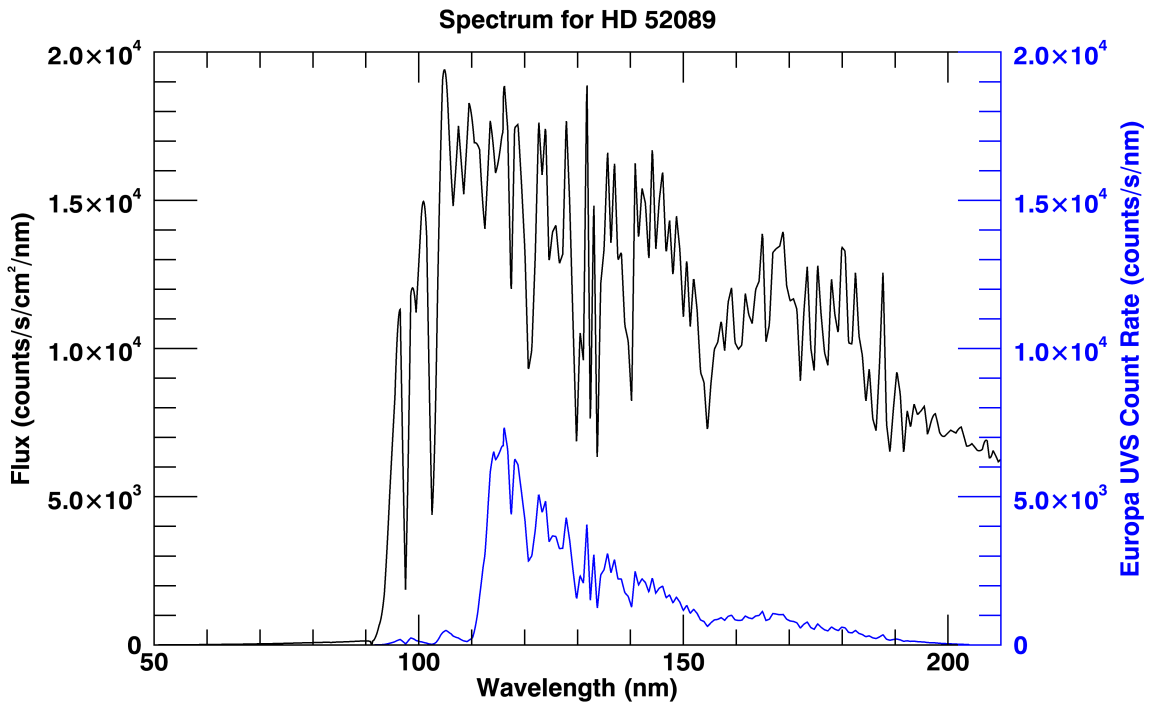

**Fig. 1** Example spectra from CUBS with both the flux of the star (black) and the estimated count rate (blue) as expected to be viewed by Europa-UVS. IUE data is used between 116-196 nm and a Kurucz model between 50-116 and 196-210 nm).

Comparing model-predicted stellar spectra with UV observations from space telescopes increases the level of confidence in these models, as demonstrated in Figure 2A. This figure shows an IUE spectrum, a Juno-UVS spectrum, and a comparison plot with Kurucz spectra. Figure 2B shows a histogram of the relative difference (RD) between the IUE and Kurucz spectra within the wavelength range of 130-175 nm. This range was chosen because it corresponds to a molecular oxygen ($O_2$) absorption feature, an important atmospheric constituent to detect on icy moons such as Europa. For example, it has also been detected in previous stellar occultations by Mars by SPICAM[23] and by the stellar appulse of Comet 67P/Churyumov–Gerasimenko[3]. These relative differences were calculated by applying Equation 1, followed byvEquation 2 to arbitrarily rank them from 0 (worst) to 10 (best).

$$RD = \left|\frac{F_{IUE}-F_{Kurucz}}{F_{IUE}+F_{Kurucz}}\right| \quad (1)$$

$$Q_{RD} = 10 + \ln\left(\frac{RD_{Max}-RD}{RD_{Max}}\right) \quad (2)$$

In the above equations, $F_{IUE}$ and $F_{Kurucz}$ are the total fluxes within the wavelength range of 130-175 nm of the IUE and Kurucz data, respectively. RD refers to the relative difference and $RD_{Max}$ is the maximum relative difference in the dataset (found to be 22). $Q_{RD}$ is the ranked quality from 0-10 which provides a simplified metric for determining whether the spectra are well-matched between the IUE catalog and the Kurucz models. Currently, those stars that have a low $Q_{RD}$ are planned to be examined on a case-by-case basis to determine the reason for the discrepancy between the model and the observed spectra.

To ensure the confidence and quality of the IUE spectra, we created a separate quality ranking metric for the Signal-to-Noise Ratio (SNR) of the IUE spectra once again within the desired wavelength range of 130-175 nm. Figure 2C shows a histogram of the SNR quality ranking for the considered stellar distribution. To get the arbitrary quality ranking from 0 (worst) to 10 (best) for this additional $Q_{SNR}$ metric we used Equation 3 where the $SNR_{Max}$ was found to be 31. Note that the bimodal distribution seen on Figure 2C is caused by the two dispersion modes available to IUE. These quality assessments are available for each of the 1800 IUE stellar targets used in CUBS, and unavailable for the remaining Kurucz-model only targets. These assessments are also currently only focused on the 130-175 nm wavelength range and thus may not be as robust for the shorter or longer wavelengths.

$$Q_{SNR} = 10 + \ln\left(\frac{SNR}{SNR_{Max}}\right) \quad (3)$$

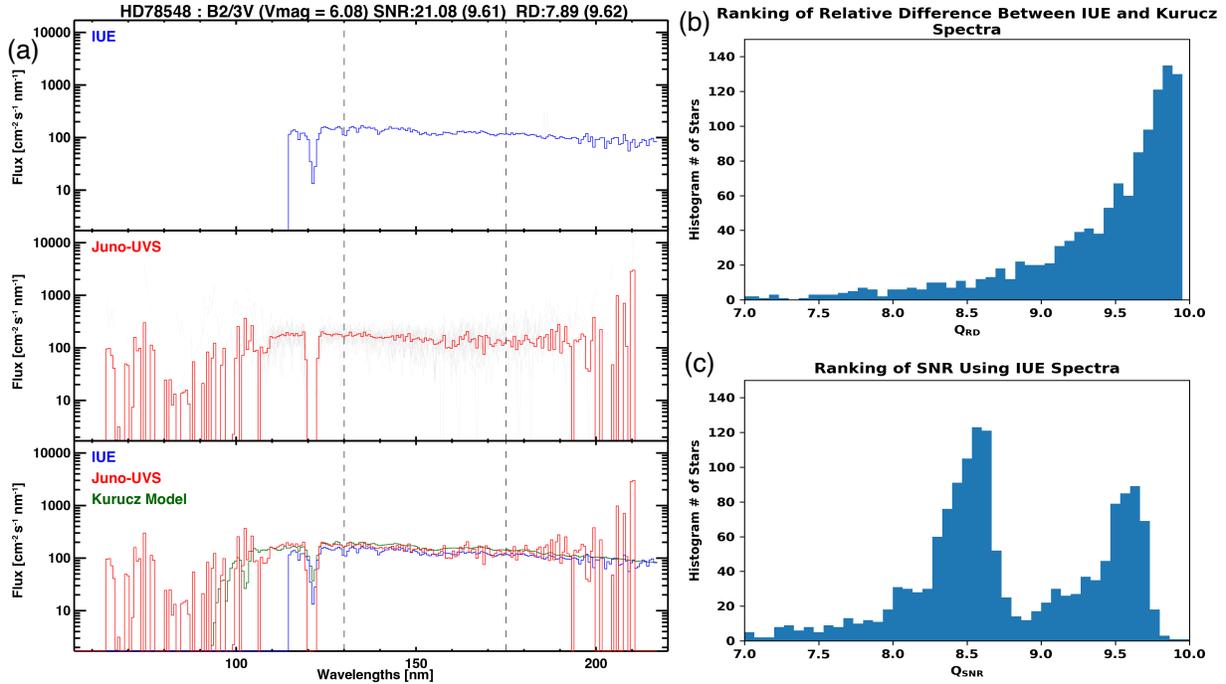

**Fig. 2** (a) Comparison of IUE (blue), Juno-UVS (red), and Kurucz (green) spectra for HD 78548. The gray lines are the total data for both IUE and Juno-UVS with the colored lines being the averages. The dashed lines are the cutoff of 130 nm to 175 nm used for the confidence factors. This is one of many examples showing how effective the Kurucz models can be to construct CUBS spectra. (b) Histogram of quality ranking [0,10] for all IUE stars in CUBS based on the relative difference (RD) between IUE and Kurucz spectra between 130-175 nm. (c) Histogram of quality ranking [0,10] for all IUE stars in CUBS based on the SNR of IUE spectra between 130-175 nm.

## 3 PLANNED APPLICATIONS

### 3.1 *Stellar occultations*

One primary planned use of CUBS is to model stellar occultation observations that will be observed by Europa-UVS and JUICE-UVS, which can provide a wealth of information relevant for atmospheric studies. A stellar occultation occurs when an observed background star passes behind a target of interest. If the star passes sufficiently close to but never behind the solid body of the target, the observation is referred to as a stellar appulse and can still provide constraints on the composition of the atmosphere above the limb. As the starlight passes through the atmosphere of the target, its observed spectrum will display absorption signatures at the relevant wavelengths of the components present in the atmosphere, which we refer to as "dips". The depth of the absorption feature ultimately indicates the abundance of that specific compound, which can be monitored until the star is completely obstructed by the solid surface. Using the absorption cross sections for each component, this data can be analyzed to provide direct information about the composition of the atmosphere as a function of altitude above the body's surface. In addition, occultation observations can be used to provide astrometric information on the topographic shape of a target based on the time of complete signal obscuration, as is planned to be done on Europa by Europa-UVS[29]. This process has been successfully implemented for constraining the sizes of asteroids in our Solar System, such as the Trojan asteroids Patroclus and Menoetius[30].

These occultation events are time-critical and need to be planned in advance. To maximize their impact, a quality factor can be established when ranking the possible choices of events to schedule. For a spacecraft such as Europa Clipper that will be orbiting Jupiter and performing close flybys of Europa, and considering that CUBS has almost 90,000 targets, there are many possible candidates for occultation measurements throughout the course of the mission. Establishing a semi-quantitative quality factor allows for an efficient way of determining which stars will be the best targets during the course of the mission.

To create the quality factor, we start by using the measured effective area for the instrument in question to convert CUBS stellar flux spectra to the instrument count rate spectra. In this work, we consider the effective area for Europa-UVS and for JUICE-UVS. Next, we isolate an on-band region that corresponds to $O_2$ absorption ranging from 130-175 nm. While $H_2O$ and several other expected species have detectable UV absorption features (Figure 3), $O_2$ is expected to be the easiest species to measure globally in Europa's atmosphere due to its relatively large abundance (ref), and hence is the best for setting a minimum criterion to determine whether an observation is useful or not. Additional observations planned for astrometric characterizations of the solid surface shape will be defined with a looser minimum criterion for its usefulness based on a complete attenuation of signal at all wavelengths. For the $O_2$ on-band range, the count rate ($R_{On}$) can be determined using the unocculted on-band rate presented in the catalog ($R_{On_0}$) which are related by the following equation:

$$r = \frac{On-band\ occultation\ Count\ Rate}{On-band\ baseline\ Count\ Rate} = \frac{R_{On}}{R_{On_0}} \quad (4)$$

which can be rearranged to:

$$R_{On} = R_{On_0} * r \quad (5)$$

where r is the transmission of the star's light through the atmosphere. Once $R_{On}$ is known we can find the total expected counts over this range ($C_{On}$) using Equation 6.

$$C_{On} = R_{On} * t_{int} \quad (6)$$

where $t_{int}$ refers to the integration time of the occultation, which is the time while the star's light is being occulted by the target's atmosphere within its scale height. This time can be found using Equation 7.

$$t_{int} = \frac{H}{v_{los}} \quad (7)$$

H represents the scale height of the atmosphere, which for Europa we set to 50 km for the general atmosphere and 30 km for a plume region. The scale height of an atmosphere is defined as the distance over which the pressure decreases by an exponential factor[31]. The velocity term $v_{los}$ is related to the angular line-of-sight velocity of the star projected to Europa's sky-plane. We can also calculate $C_{On_0}$ (Equation 8), which represents the total counts in the on-band wavelength region of the unocculted star to serve as a baseline for the lightcurve (i.e., the measured signal as a function of time).

$$C_{On_0} = R_{On_0} * t_{baseline} \quad (8)$$

The baseline time ($t_{baseline}$) to get these initial counts is meant to be time spent above 400 km at Europa. Nominally, the goal is to get 10 minutes of baseline time, but any time period that achieves an SNR > 3 for the lightcurve as a whole is sufficient. Once the counts are determined, an expected dip in signal can be determined based on a canonical example modeled atmosphere as explained later in this section. This dip in the signal is calculated using Equation 9

$$d = 1 - r \tag{9}$$

where d refers to the aforementioned absorption feature "dip" in the lightcurve during the occultation. The dip then goes through a propagation of error to determine its uncertainty as shown in Equation 10.

$$\sigma_d = \sigma_r = r\sqrt{(\frac{\sigma_{R_{On}}}{R_{On}})^2 + (\frac{\sigma_{R_{On_0}}}{R_{On_0}})^2} = r\sqrt{(\frac{\sqrt{C_{On}}}{C_{On}})^2 + (\frac{\sqrt{C_{On_0}}}{C_{On_0}})^2} = r\sqrt{\frac{1}{C_{On}} + \frac{1}{C_{On_0}}} \tag{10}$$

The uncertainty found in Equation 8 ($\sigma_d$) is then used to calculate the dip's SNR as given in Equation 11.

$$SNR_d = \frac{d}{\sigma_d} = \frac{d}{r\sqrt{\frac{1}{C_{On}} + \frac{1}{C_{On,0}}}} \tag{11}$$

In total, we have 3 different regimes for absorption criteria that depend on the target science objective. The first is the case of a general atmosphere where we set the $O_2$ absorption feature to be a 10% drop in the count rate and we use 50 km for the scale height. The second case is a plume dominated region where the absorption is set to be a 50% dip and the scale height is 30 km to represent the altitude range where the plume's ejecta will likely dominate. The last case is full obstruction behind the moon, where there would be no transmission of light; this will be used for astrometry. In this case, the scale height is set to 100 m and the count rate is measured across the entire wavelength range, not just the on-band because it is not composition-dependent. Figure 3 shows an example of absorption through Europa's atmosphere that uses a more sophisticated model than needed for this initial quality factor that can be employed once a narrowed list of candidates has been established using our quality factor.

Other signal in the spectra include Europa disk reflectance as well as airglow emissions and radiation-induced detector noise which is straightforward to subtract in our signal but adds additional uncertainties. For stars brighter than Europa's disk reflectance, this noise correction will be largely ignorable. For example,

$$\sigma_{C_{On}} = \sqrt{\frac{1}{C_{On}} + Noise^2} \tag{12}$$

$SNR_d$ gives a quality factor that helps rank the subjective quality of the stellar occultation as "Very Good", "Good", "Ok", or "Bad". These rankings correspond to SNRs of >100, 10-100, 3-10, and <3, respectively. Once the categories are established using the semi-quantitative quality factor, additional subjective criteria can also be employed to move an occultation up or down in ranking. These criteria are dependent on the objectives and specifics of the particular mission. For example, Europa-UVS has the criteria listed below which are subject to change as the mission progresses.

- Fills global location and time coverage needs
- Passes through candidate plume locations

- Previously observed with Europa-UVS
- Has IUE or other known spectra, not only models
- Low flux variability and/or high accuracy of parameters such as astrometry
- Emphasizes extreme-UV (< 115 nm) capability and sensitivity to additional species
- Pair of events, two cords at once (lines connecting ingress and egress for astrometric calculations), or proximity to another UV-bright star
- Times when background radiation levels near the spacecraft are low
- Follows the Clipper trajectory line of sight for best in-situ (MASPEX + SUDA) comparisons
- Lines up with radar (REASON) groundtrack for best astrometry correlation
- Coincides with images (EIS) of high phase limb plume search locations
- Relatively less auroral brightness, impacting cases with borderline SNR
- Near-simultaneity with JUICE encounters
- Near-simultaneity with JUICE-UVS occultation observations

This process is currently being utilized by the Europa-UVS planning team and can be replicated for any other UV mission/instrument (including plans for JUICE-UVS) given knowledge of the instrument effective area and defined set of spacecraft trajectories during the mission lifetime.

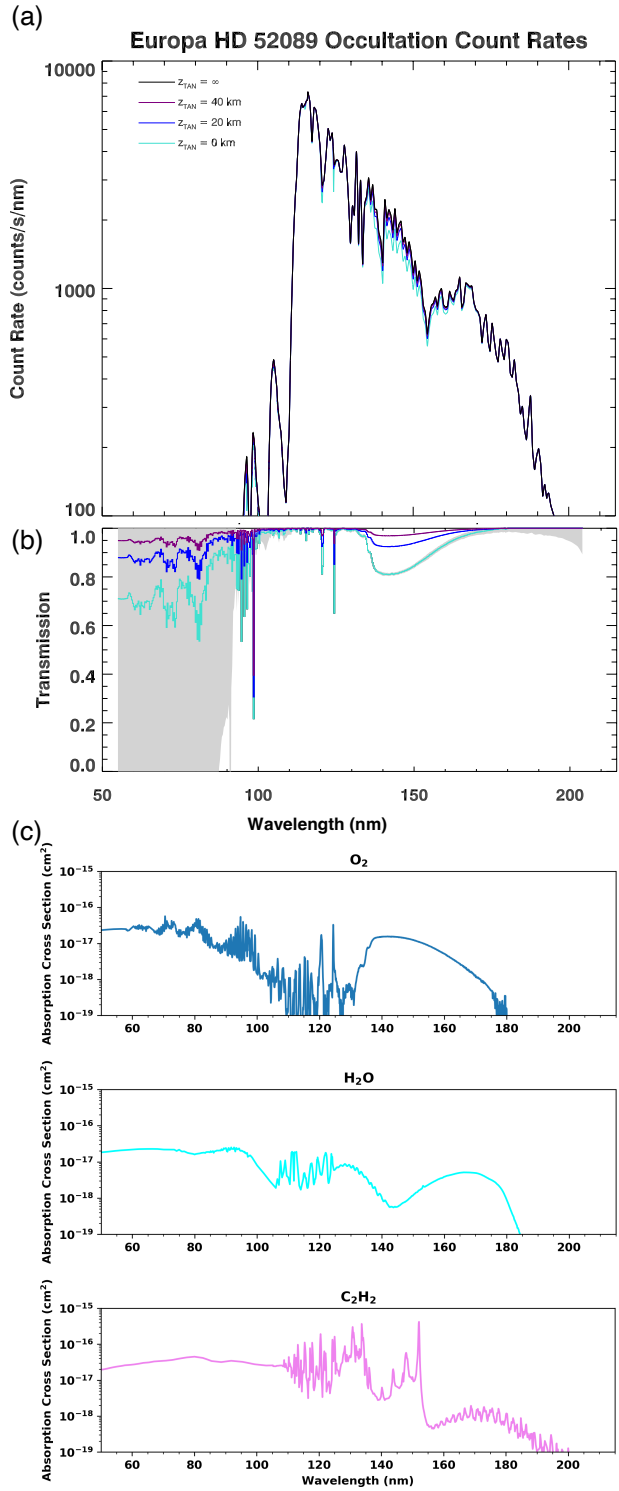

**Fig. 3** Example (a) stellar occultation spectra and (b) transmission spectra from star HD 52089 showing how the stellar signal decreases in intensity at specific wavelengths related to the composition of a model for Europa's atmosphere as the star gets closer to the surface (altitude $Z_{TAN}$ = 0 km) from the reference frame of the observing spacecraft. The gray region represents the $1\sigma$ uncertainty for the transmission spectra. (c) Absorption cross sections for key species in Europa's atmosphere ($O_2$, $H_2O$, and $C_2H_2$) used in the present model.

**3.2** *Starlight illumination modeling*

The next intended use of CUBS involves starlight illumination modeling, which allows for surface reflectance studies of various airless Solar System bodies. This activity is performed by using a UV spectrograph to observe a target's nightside surface when there is no background light from the Sun. As a result, the only light shining on the surface is that from UV-bright stars and other background sources (background sky, Jupiter shine, satellite shine), and so any reflectance observed by the spectrograph is a result of all of these diffuse background illumination sources. The reflectance can then be analyzed at specific wavelengths to characterize the surface composition and microphysical structure (e.g., compaction/porosity), as for dayside sunlit imaging.

This approach was demonstrated using an earlier version of the catalog at the Moon to study its permanently shadowed regions (PSRs) and on the lunar nightside using the LAMP instrument on NASA's LRO[10,11]. Byron et al.[11] expanded the initial IUE-based catalog using Kurucz models and added the Large and Small Magellanic clouds as UV illumination sources, based on Juno-UVS observations of these extended sources[13]. They used LRO LAMP observations during nighttime to create brightness maps and then divided by the incident stellar illumination predicted by the catalog to create a far-UV albedo map of the Amundsen crater on the Moon. Their results from these observations were important in determining that the albedo of the PSR inside the crater is lower than the surrounding regions, allowing them to estimate porosities of the regolith. The catalog in its current, further-expanded state allows for a more accurate representation of the background stars contributing to the illumination of nightside surfaces. We plan to implement CUBS when modeling the stellar, interplanetary Lyman-alpha and helium skyglow, and Jupiter-shine illumination of the nightside surfaces of Europa, Ganymede, and Callisto as viewed by Europa-UVS and JUICE-UVS.

**3.3** *Galactic UV Sky Map*

Another intended use of CUBS involves studying diffuse galactic UV dust emissions. By plotting each star by location using their RA and dec along with their fluxes, we can make a sky map in the UV. We specifically integrated first from 115-210 nm to cover the far-UV (FUV) range and next from 50-115 nm to cover the extreme-UV (EUV) range (Figure 4). After applying some smoothing along the galactic plane, we directly compared the maps derived from the spectral models to maps generated by Juno-UVS[12] and other UV instruments. Through comparisons between observations and our modeled sky, we find that CUBS serves as a reliable representation of the stars present in the galaxy and their fluxes. However, there is an apparent glow, especially along the galactic plane, that is observed by Juno-UVS but cannot be completely modeled with just the included stars, indicating that additional UV sources (for example, dust-scattered light from background UV stars) is present. This unmodeled glow requires further study and CUBS will be useful for separating the dust-glow from the stellar components. A similar photometric simulation based on IUE catalog sources has been applied to the GALEX dataset for similarly isolating the non-stellar sources[32].

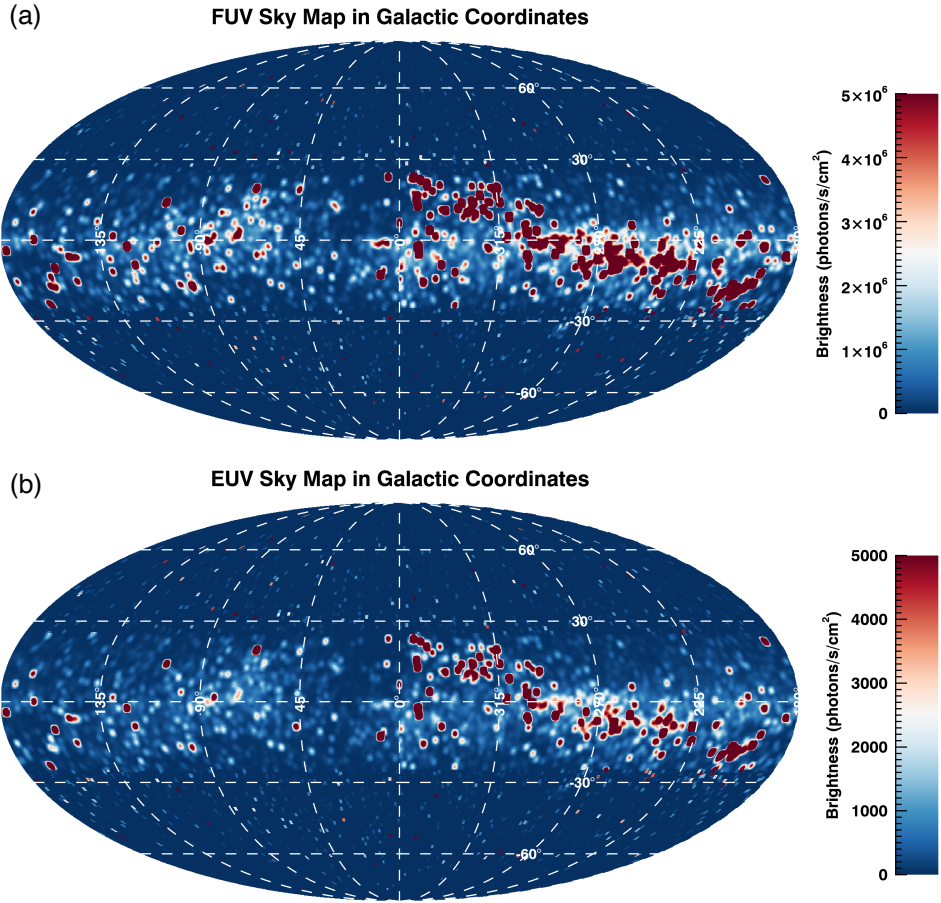

**Fig. 4** (a) FUV skymap created using CUBS integrated from 115-210 nm. (b) EUV skymap integrated from 50-115 nm.

## 4 CONCLUSIONS

A robust UV stellar catalog, such as the one presented here, can serve as an invaluable resource to those performing UV spectroscopic studies that utilize UV-bright stars. CUBS was created using a combination of Kurucz models and IUE spectra, where available. For those stars with available IUE spectra, we compared the Kurucz and the IUE spectra to assess the efficacy of the Kurucz models. We also compared these spectra to those available from Juno-UVS calibration to further provide a confidence test for the Kurucz models. We found that there is a strong agreement between them, which demonstrates that the models are sufficiently reliable for the purposes of observation planning. This paper also discussed three possible applications of CUBS:

1. Planning and simulating stellar occultations
2. Modeling illumination of dark, nightside planetary surfaces primarily lit by the background sky
3. Investigating the origin of diffuse galactic UV light as hinted by previous UV datasets

Stellar occultations are a key tool for studying atmospheric compositions and structure throughout the solar system. Starlight illumination modeling is important for analyzing surface composition in otherwise dark regions on planetary surfaces. Future UV sky map investigations are needed to understand the full extent of the diffuse components that contribute to the galaxy's glow. Several planned uses were described for supporting the upcoming Europa Clipper and JUICE mission UV investigations.

**Acknowledgements**

This work expands upon an SPIE Proceedings paper submitted for the 2022 SPIE Astronomical Telescopes and Instrumentation Conference[33] and was supported by NASA through the Europa Clipper Project.

**Michael Velez** is beginning his fourth year of enrollment in the Southwest Research Institute/University of Texas at San Antonio Joint Graduate Program in Space Physics and Instrumentation. Michael also serves as a Graduate Affiliate on the Europa Clipper Mission. His current topics of research include developing a stellar catalog for ultraviolet research, analyzing optical emissions at Europa using the Hubble Space Telescope (HST), and constraining molecular hydrogen levels at Europa using HST far-ultraviolet data.